\documentclass[prc,aps,preprint,showpacs]{revtex4}
\usepackage{graphicx}

\begin{document}
\title{Pigmy resonance in monopole response of neutron-rich 
Ni isotopes ?}

\author{ Ikuko Hamamoto$^{1,2}$ and Hiroyuki Sagawa$^{1,3}$}

\affiliation{
$^{1}$ {\it Riken Nishina Center, Wako, Saitama 351-0198, Japan } \\ 
$^{2}$ {\it Division of Mathematical Physics, Lund Institute of Technology 
at the University of Lund, Lund, Sweden}  \\
$^{3}$ {\it Center for Mathematics and Physics, University of Aizu, 
Aizu-Wakamatsu, Fukushima 965-8560, Japan} } 




\begin{abstract}

The RPA monopole strength as well as the unperturbed particle-hole 
excitation strength is studied, in which the strength in the continuum 
is properly treated without discretizing unbound particle spectra.  
The model is the sef-consistent Hartree-Fock calculation and the RPA Green's
function method with Skyrme interactions.  
Numerical examples are the Ni-isotopes, especially $^{68}_{28}$Ni$_{40}$, 
in which an experimental observation 
of low-lying socalled ''pigmy resonance'' with an appreciable amount of 
monopole strength at 12.9 $\pm$ 1.0 MeV was recently reported.   
In the present study it is concluded that sharp monopole peaks 
with the width of the
order of 1 MeV can hardly be expected for 
$^{68}$Ni in that energy region.  Instead, a broad shoulder 
of monopole strength consisting of neutron excitations to   
non-resonant one-particle states  
(called ''threshold strength'') with relatively 
low angular-momenta $(\ell, j)$  
is obtained in the continuum energy region above the particle threshold, 
which is considerably lower than that of 
isoscalar giant monopole resonance. 
In the case of monopole excitations of $^{68}$Ni  
there are no unperturbed particle-hole states below 20 MeV, in which the
particle expresses a neutron (or proton) resonant state.    
It is emphasized that in the theoretical estimate a proper treatment of 
the continuum is extremely important.

\end{abstract}


\maketitle

\newpage

Using inelastic alpha scattering at 50$A$ MeV on the unstable nucleus 
$^{68}$Ni in inverse kinematics M. 
Vandebrouck {\it et al.} recently observed \cite{MV14} a 
soft monopole mode at 12.9 $\pm$ 1.0 MeV, in 
addition to the isoscalar 
giant monopole resonance (ISGMR), of which the centroid is placed 
at 21.1 $\pm$ 1.9 MeV.  This obsevation is the first one, in which 
the low-lying socalled pigmy resonance of monopole type is reported.
The pigmy resonance reported in Ref. \cite{MV14} was said to be approximately in 
agreement with the result of the calculation of Ref. 
\cite{EK11}, in which the random phase approximation (RPA) calculation 
is performed based on the discretized particle spectra. 
First of all, it is our impression that though the experimental result reported  
in Ref. \cite{MV14} is very exciting, the observation of the pigmy resonance in
the monopole response function must be further confirmed by additional 
experiments in the future.    
This is because the angular distribution of the 12.9 MeV peak 
for the forward angle, 
$\theta_{CM} < 5^{\circ}$, is not reported, which is needed 
to identify exclusively the monopole strength.  
Secondly, in the case of monopole excitations of $^{68}$Ni 
it is difficult for us to imagine the unperturbed 
discrete particle-hole (p-h) excitations below 20 MeV, which can be properly   
used as the basis for the
RPA calculation.     
Consequently, it is hard to expect that some  
monopole peaks with the widths of the order of 1 MeV appear in the 
low-energy region.
The RPA calculation, which is performed using positive-energy particle 
states obtained by discretizing
particle spectra in the continuum, is not acceptable, because 
the energies as well as wave functions of such particle states have, 
in general, no correspondence to those of 
one-particle resonant states.

In halo nuclei, in which the wave function of very weakly 
bound $\ell$ = 0 or 1 nucleon(s) extends largely to the outside of the core
nucleus, the threshold strength that comes from the excitation of 
such bound nucleon(s) to    
non-resonant one-particle states, 
may appear as a relatively 
sharp peak slightly above the particle threshold \cite{SAF91}. 
In contrast, in nuclei such as $^{68}$Ni, of which 
the neutron separation energy is 7.8 MeV, the shoulder 
originating from the threshold strength is generally broad and the 
width may be 
the order of several MeV or larger. Consequently, the strength may be
difficult to be experimentally separated from the background strength.

It may sometimes occur especially for neutrons 
with small orbital angular-momenta 
$\ell$ that no one-particle resonant states are available, to which those
neutrons are excited by a given low-multipole excitation operator.  
In such cases
the entire p-h strength for the neutrons in a given hole orbit 
would appear as threshold strength \cite{HSZ98a}.    
One such example is the isoscalar (compression) dipole strength of the
neutron-drip-line nucleus $^{22}$C \cite{HSZ98}, in which the major strength
should appear as the threshold strength with a low-energy broad peak 
instead of an easily recognizable giant resonance in a high-energy region.      
The only way to reliably calculate the response functions in such cases is to
treat the continuum as it is.  

More than 15 years ago we have studied response functions of 
various type (isoscalar and isovector multipole modes, charge 
exchange modes and spin-dependent modes) in unstable 
nuclei, using the 
Hartree-Fock (HF) plus the self-consistent 
RPA with Skyrme interactions, which is solved in coordinate space using the 
Green's function method.
One of the characteristic features of the response functions in 
neutron-rich nuclei is the possible appearance of an appreciable amount of 
low-energy neutron threshold 
strength, which is not related to a resonant behavior of 
unbound single-particle states. 
For a given bound 
neutron with the binding energy $\varepsilon_B$ 
the relevant threshold strength starts to appear 
at the excitation energy ($E_x$) equal to 
$\varepsilon_B$.  The strength of excitations to neutron orbits 
with $\ell$ rises 
as the $\ell + \frac{1}{2}$ power of available energies, 
$(E_x - \varepsilon_B)^{\ell + 1/2}$ \cite{BW52}, which   
rises more steeply for lower $\ell$-values just above  
the threshold energy \cite{SAF91, HS95}. The shape of the threshold strength is
discussed in, for example, Ref. \cite{NLV05}.  
One of the well-known examples of the threshold strength is the sharp and 
strong low-energy peak of the dipole strength in halo nuclei, 
which is detected, for example,  
by Coulomb break-up reactions of halo nuclei \cite{TN94}.  

Since the model used in the present paper is  
the same as that 
used in our paper in 1997 \cite{HSZ97}, here we write only the necessary 
points. 
Neglecting the pair correlation, 
the self-consistent HF+RPA calcurations with the Skyrme
interactions are performed in coordinate system, and the strength distributions
$S(E)$ are obtained from the imaginary part of the RPA Green function,
$G_{RPA}$, as 
\begin{equation}
S(E) \: = \: \sum_{n} \mid <n\mid Q \mid 0> \mid ^{2} \delta (E-E_{n})
\: = \: \frac{1}{\pi}\, Im\, Tr (Q^{\dagger } (\vec{r})\, 
G_{RPA}(\vec{r};\vec{r'};E)\, Q(\vec{r'})) \qquad . 
\label{eq:SE}
\end{equation}
where  ~$Q$~ expresses one-body operators 
\begin{equation}
Q^{\lambda=0, \, \tau=0} \: = \: \frac{1}{\sqrt{4 \pi}} \, \sum_{i} r^{2}_{i} 
\qquad \mbox{for isoscalar monopole strength.}
\end{equation}

In Fig. 1 the monopole strength of the nucleus $^{68}_{28}$Ni$_{40}$  is shown,
which is calculated using the SLy4 interaction.  The neutron separation energy
$S_n$ is experimentally 7.79 MeV, while 8.92 MeV for the SLy4 interaction,
which is equal to the binding energy of the occupied least-bound neutrons 
in the $1f_{5/2}$ orbit.  The calculated energies of neutrons in $2p_{1/2}$, 
$2p_{3/2}$, and $1f_{7/2}$ orbits are $-$9.25, $-$11.39, and $-$16.45 MeV, 
respectively.  The proton separation energy $S_p$ is experimentally 15.43
MeV, while 14.9 MeV for the SLy4 interaction, which is equal to the binding
energy of the occupied least-bound protons in the $1f_{7/2}$ orbit.  
In Fig. 1(a) 
it is seen that the RPA strength increases monotonically from the
particle threshold to the ISGMR peak  
around 21 MeV, in contrast to Fig. 1 of Ref. \cite{EK11}.  
It is also seen that the low-energy threshold strength is basically of
single-particle nature and obtains only a minor influence by particle-hole
correlations.    
In Fig. 1b the neutron unperturbed monopole strengths, which contribute
appreciably to the total unpertubed strength below 20 MeV shown in Fig. 1a, 
are shown for respective occupied hole orbits.  
In the monopole case the $(\ell, j)$-value of 
particle states must be the same as that of respective hole orbits, and   
none of the neutron one-particle resonances with  
$3p_{1/2}$, $2f_{5/2}$, $3p_{3/2}$, and $2f_{7/2}$
exist.  Consequently, the entire monopole strengths related to 
those neutrons in the hole orbits appear  
as respective threshold strengths in Fig. 1(b).    

No smearing
is made in our calculated strength in Fig. 1, while smearing is
already made in Fig. 1a of Ref. \cite{EK11}. 
The reason why the ''pigmy resonance'' is obtained 
around 15 MeV in Ref. \cite{EK11} is explained by the sentence in
\cite{EK11}, 
''Because of the neutron excess in $^{68}$Ni, 
the $3p2f$ states are located close
to the Fermi level and, moreover, the $2p1f$ shell is calculated just below the
Fermi level''.  
In order to make the above sentence meaningful, 
we would naturally interpret that their $3p2f$ states are supposed to 
approximately represent 
respective one-particle resonant states.  However, we find no
such $3p2f$ one-particle resonant states in the continuum spectra.    

If pair correlation is negligible and one discretizes 
the unbound particle spectra by using an  
available discretization method, such as confining the system in a finite 
box or expanding the wave-functions in terms of a finite number of 
harmonic-oscillator basis \cite{EY13,DV12},  
one may obtain, for example, only one discrete state with a given $(\ell j)$ 
in the
energy range of 5 MeV above the threshold of respective $(\ell j)$ particles.    
When the size of the finite box increases or the number of basis increases, 
the number of the given ($\ell j$) particle states 
in the same energy range increases and the energy of the
lowest-lying discretized state decreases.   
As the number of basis approaches infinite, the shape of the total strength 
coming from 
many discretized particle states with the given $(\ell j)$ may approach the
shape of the curve shown in Fig.1b.    
The consideration above explains that the lowest-lying 
unbound particle states obtained
by discretizing the continuum with an easily manageable finite number of basis 
can be neither similar to one-particle resonant
states nor the proper representatives of threshold strength.  
In short, a great caution must be taken against using 
discretized particle states 
as the constituents of continuum particle states in RPA.  

Using the definition of one-particle resonant energy, at which the phase shift
increases through $\pi / 2$ as the energy increases, for the present HF
potential of $^{68}$Ni with the SLy4 interaction we obtain one-particle energy
of the $2d_{5/2}$ neutron at $-$0.71 MeV and one-particle resonant energy of the
$2d_{3/2}$ neutron at +1.13 MeV (with the width of 0.54 MeV). Thus, the
calculated spin-orbit splitting for $2d$ neutrons is 1.13 $-$ ($-$0.71) = 1.84
MeV.  Similarly, one-particle resonant energies of the $2d_{5/2}$ and 
$2d_{3/2}$ protons are obtained at +1.27 and +3.23 MeV, respectively, with
negligible widths.  Thus, the calculated spin-orbit splitting for $2d$ protons
is 3.23 $-$ 1.27 = 1.96 MeV.  As seen from these numerical examples,  
spin-orbit splittings of one-particle resonant levels do not approach zero,
though they are reduced
compared with spin-orbit splittings of deeply-bound one-particle states.  In
contrast, the following sentence is found in Ref. \cite{EY13}, ''... This
expected vanishing of the spin-orbit splitting for unbound states ...'', which
was made based on the discretized continuum particle spectra.  
We are afraid that ''unbound states'' in the sentence correspond to 
some of the infinite number of continuum states.  
Furthermore, in
Ref. \cite{MV14} we find a statement, ''the
observation of the soft monopole mode could bring valuable information on
spin-orbit splitting.'', which was presumably made based on the numerical
results in Ref. \cite{EY13}.    
It is difficult for us to find a sensible meaning in
the statement because of the non-resonant nature of the monopole strength below
the ISGMR. 

In Fig. 2 the monopole strength of the nucleus $^{78}_{28}$Ni$_{50}$  is shown,
which is calculated using the SLy4 interaction.  The neutron separation energy
is not known exprimentally, while 5.88 MeV for the SLy4 interaction,
which is equal to the binding energy of the occupied least-bound neutrons 
in the $1g_{9/2}$ orbit.  The calculated energies of neutrons in $2p_{1/2}$, 
$1f_{5/2}$,  $2p_{3/2}$, and $1f_{7/2}$ orbits are 
$-$10.11, $-$10.71, $-$12.04 and $-$17.48 MeV, 
respectively.  The proton separation energy is not known experimentally, 
while 20.8 MeV for the SLy4 interaction, which is equal to the binding
energy of the occupied least-bound protons in the $1f_{7/2}$ orbit.  
In Fig. 2a one sees that the 
properties of ISGMR of $^{78}$Ni are similar to those of $^{68}$Ni.  
The main difference between the monopole strengths of $^{68}$Ni and $^{78}$Ni
is: due to the extra filling of the neutron $1g_{9/2}$ orbit in $^{78}$Ni, 
$S_n$ is considerably smaller, $S_p$ is larger, and the height of the
broad flat shoulder of the neutron threshold strength is appreciably higher. 
The RPA strength increases monotonically from the
particle threshold to the ISGMR peak  
around 21 MeV.  The strength function in Fig. 2a, which is not smeared out, may
be compared with Fig. 5a in Ref. \cite{EK11} which is the result of smearing
out.    
In Fig. 2b the neutron unperturbed monopole strengths, which contribute
appreciably to the total unpertubed strength below 20 MeV in Fig. 2a, 
are shown for respective occupied hole orbits.    
Since none of the neutron one-particle resonances exist for the   
$2g_{9/2}$, $3p_{1/2}$, $2f_{5/2}$, $3p_{3/2}$, and $2f_{7/2}$ orbits
the entire monopole strengths related to 
those neutrons appear  
as the respective threshold strengths.    

In conclusion, 
based on our continuum calculation without discretizing continuum particle
spectra, 
we state that 
it is very unlikely to have some isoscalar monopole peaks with the width 
of the order of 1 MeV below the excitation energy of 20 MeV in $^{68}$Ni.  
In nuclei other than halo nuclei 
the low-lying isoscalar monopole threshold strength 
may appear as a broad shoulder as in the present
neutron-rich Ni-isotopes. The broad shoulder may sometimes look like 
a broad bump.  An example is 
the calculated isoscalar monopole strength 
in $^{60}$Ca shown in Fig. 12 of Ref. \cite{HSZ97}. However, the
''width'' of the bump in such cases is several MeV or larger.  
In order to compare our results with the observed strength, 
we may further smear out our RPA curve so as to approximately include 
the effect of 2p-2h components that are not taken care of in RPA.  
Then, the ''width'' of the bump would become even larger.

\newpage

\noindent
{\bf\large Figure captions}\\
\begin{description}
\item[{\rm Figure 1 :}]
(Color online) Monopole strength function (\ref{eq:SE}) of $^{68}$Ni 
as a function of excitation
energy.  The SLy4 interaction is used consistently both in the HF and the RPA
calculations.  (a) Unperturbed (neutron plus proton) 
monopole strength and isoscalar monopole RPA strength. 
The RPA strength denoted by the solid curve includes 
all possible  
strengths due to the coupling between bound and unbound states in RPA.  
On the other hand, 
in the unperturbed response which is denoted by the short-dashed   
curve the p-h strengths, in which both particle and hole orbits are bound, are 
not included because of the different dimension of the strength.  
The energies of those unperturbed p-h excitations 
are the $1d_{5/2} \rightarrow 2d_{5/2}$ excitation at 27.60 MeV for neutrons and 
the excitations of $1p_{3/2} \rightarrow 2p_{3/2}$  at 27.58 MeV and 
$1p_{1/2} \rightarrow 2p_{1/2}$
at 27.46 MeV for protons.  In addition, the proton excitation at 27.3 MeV  
from the bound $1d_{5/2}$ orbit to the
one-particle resonant $2d_{5/2}$ orbit has such a narrow width 
that the
strength is not plotted.  The narrow peaks at 24.1 and 24.7 MeV in the
unperturbed strength curve are the proton $2s_{1/2} \rightarrow s_{1/2}$ and 
$1d_{3/2} \rightarrow 2d_{3/2}$ excitations, respectively.
ISGMR peak is obtained at slightly lower than 21 MeV;
(b) Some unperturbed neutron threshold strengths, 
which contribute appreciably to the
total unperturbed strength below the energy of ISGMR in Fig. 1a, 
are shown for respective occupied hole orbits, 
$(1f_{7/2})^{-1}$, $(2p_{3/2})^{-1}$, 
$(1f_{5/2})^{-1}$, and $(2p_{1/2})^{-1}$.  
Below 15 MeV the sum of the four curves
shown is exactly equal to the short-dashed curve in Fig. 1a.  
\end{description}

\begin{description}
\item[{\rm Figure 2 :}]
(Color online) Monopole strength function of $^{78}$Ni as a function 
of excitation 
energy.  The SLy4 interaction is consistently used both in the HF and the RPA
calculations.   
(a) Unperturbed (neutron plus proton) monopole strength 
and isoscalar monopole RPA strength. 
The RPA strength includes all possible strengths. 
On the other hand, in the unperturbed strength that is
denoted by the short-dashed curve the p-h strengths, in which both particle and
hole orbits are bound, are not included because of the different dimension of
the strength;   
(b) Some unperturbed neutron threshold strengths, which contribute appreciably
to the total unperturbed strength below the energy of ISGMR 
in Fig. 2a, are shown
for respective occupied hole orbits.  

\end{description}

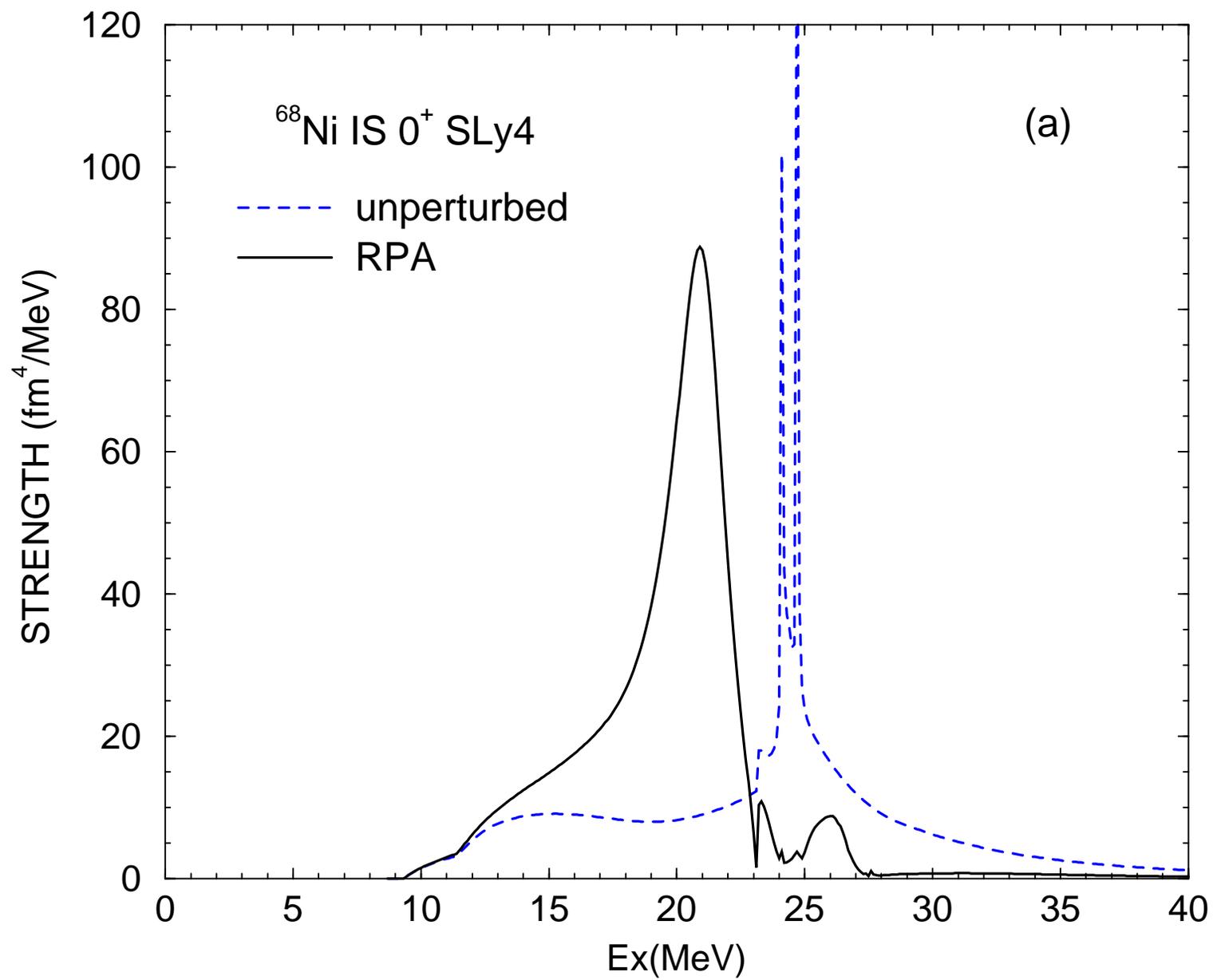
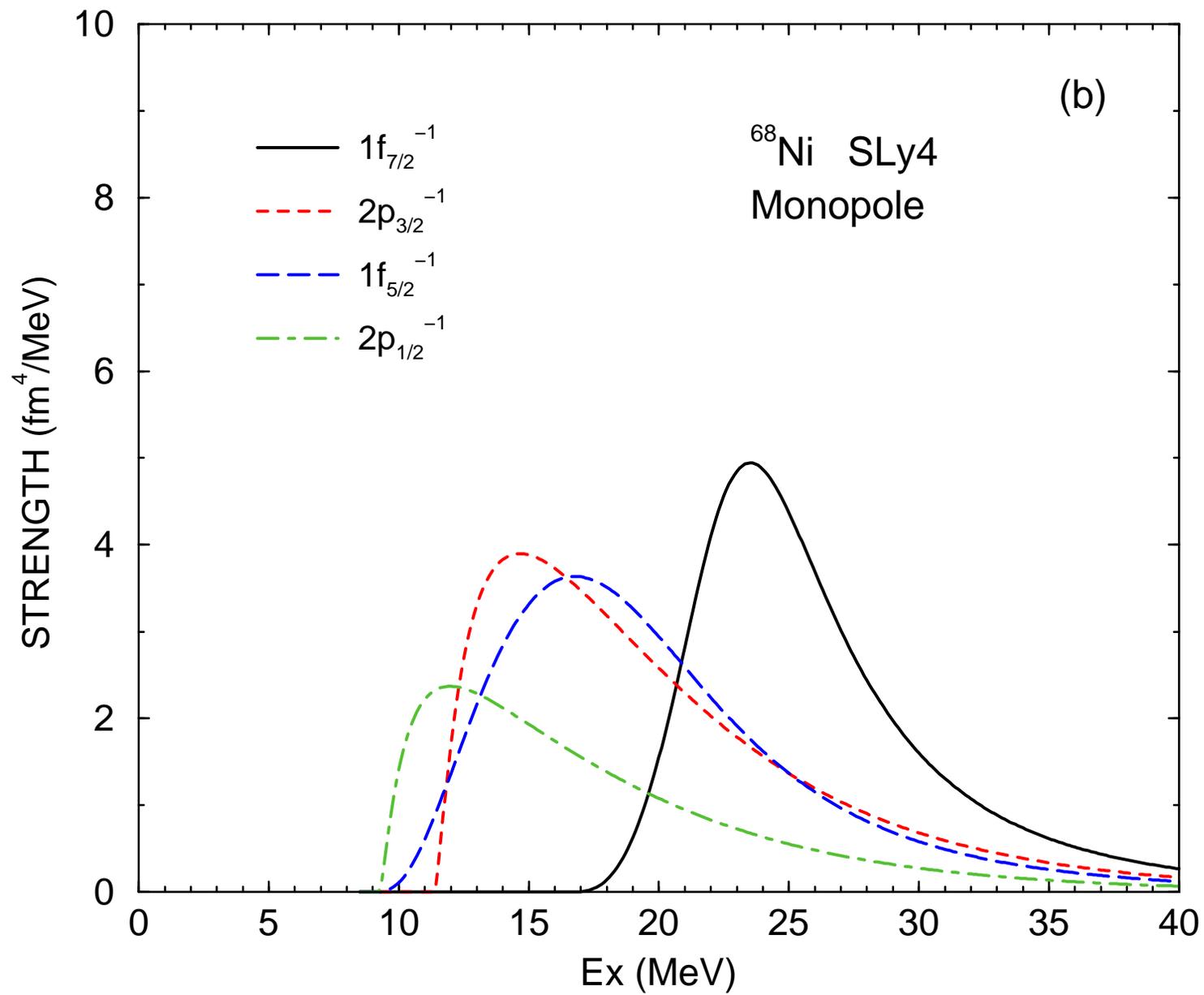
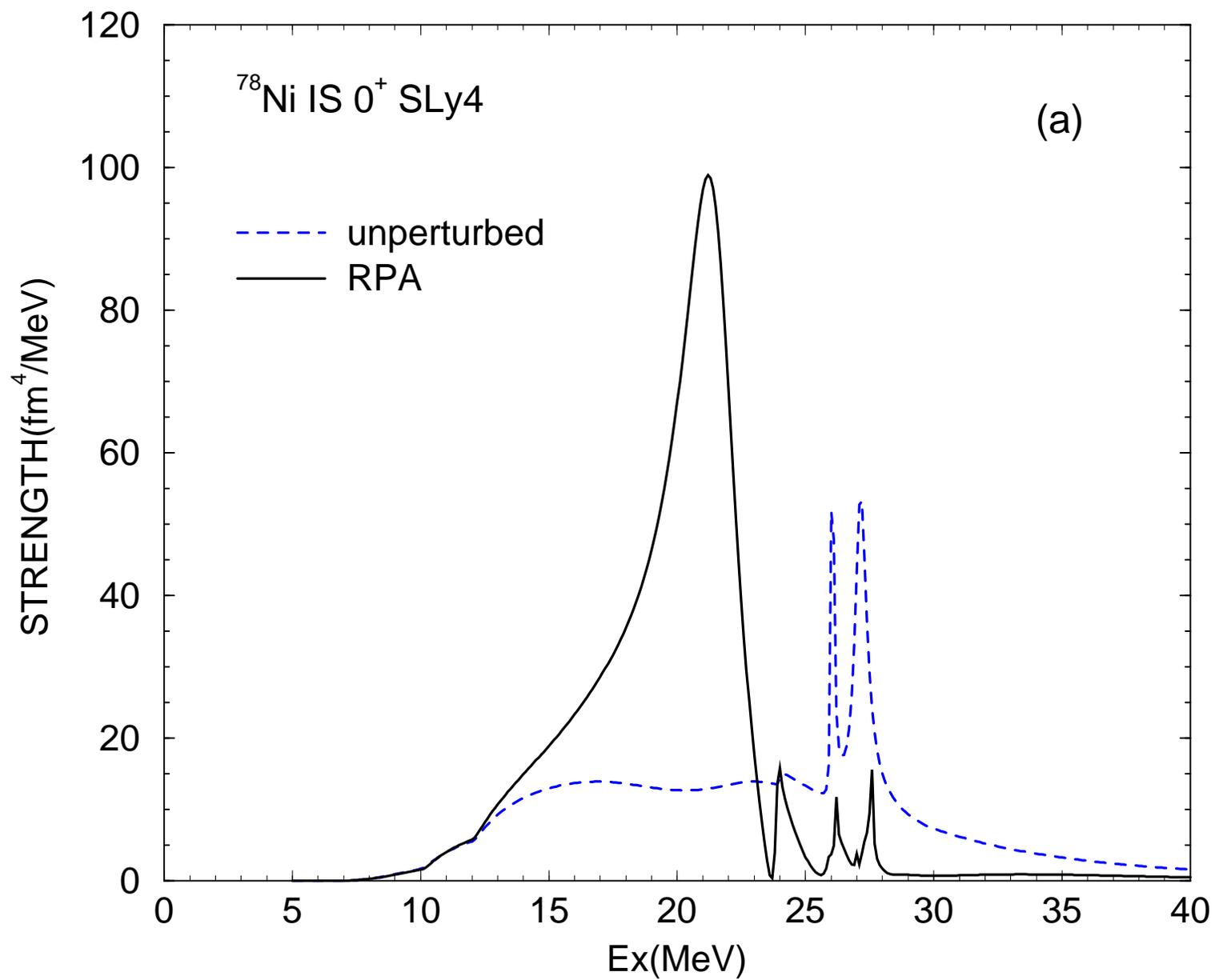
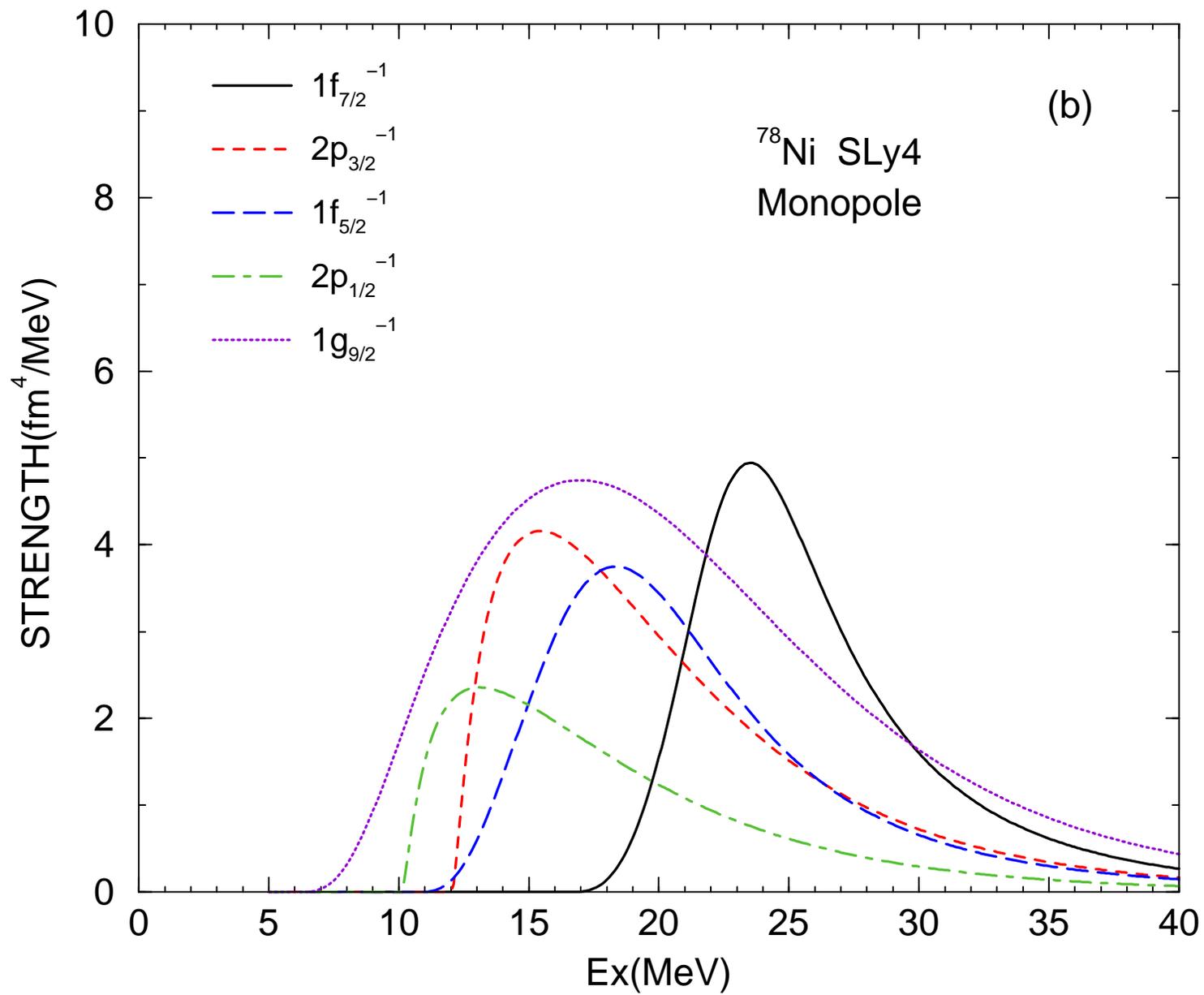
\end{document}